\documentclass{article}
\usepackage{graphicx}
\usepackage{orcidlink}
\usepackage{siunitx}
\usepackage{url}
\DeclareUrlCommand\code{\urlstyle{tt}}
\DeclareUrlCommand\path{\urlstyle{tt}}
\DeclareUrlCommand\stamp{\urlstyle{tt}}
\usepackage{array,tabularx,booktabs}
\usepackage{float}
\usepackage{tikz}
\usetikzlibrary{arrows.meta,positioning}
\begin{document}

\title{Smart Contracts Claimed Vulnerable by the CVE Database, with Labels and Source Locations}

\author{%
    Monika di Angelo~\orcidlink{0000-0002-4217-4530} $^{1,3}$ and Gernot Salzer~\orcidlink{0000-0002-8950-1551} $^{2,3}$\\
    \normalsize $^{1}$Institute of Computer Engineering, Informatics, TU Wien, Vienna, Austria\\
    \normalsize $^{2}$Institute of Logic and Computation, Informatics, TU Wien, Vienna, Austria\\
    \normalsize $^{3}$Division of Theoretical Computer Science,\\
    \normalsize KTH Royal Institute of Technology, Stockholm, Sweden\\
    \normalsize \{monika.di.angelo, gernot.salzer\}@tuwien.ac.at\\
}
\date{} 
\maketitle
\begin{abstract}
  The Common Vulnerabilities and Exposures (CVE) database catalogs vulnerability claims in hard- and software, among them those pertaining to blockchain programs a.k.a.~smart contracts.
  We present CVE-Smart-Contracts, a curated dataset of CVE records up to July 2026 referring to Ethereum smart contracts.
  The dataset contains the vulnerable artifacts (source code and runtime bytecode), labels according to three taxonomies, and function-level locations.
  The retrieval of CVE records, collection of additional evidence, validation of the correspondence between records and artifacts, label assignment, and vulnerability localization are automated, leaving 15\,\% to manual analysis.
  The dataset does not validate the original vulnerability claims, but marks a few records obviously wrong as `refuted'.
  For the sake of reproducibility, all external inputs are retained, so that rerunning the pipelines results in the same outputs.
  The dataset comprises 491 records linked to deployed contracts, 26 referring to projects (mostly libraries), 45 without validated artifacts, and six records with refuted claims.
  The dataset supports empirical security research, in particular the evaluation of code analysis and repair techniques.
\end{abstract}

\section*{Background \& Summary}

When developing and evaluating vulnerability detection techniques, there is a constant need for reliable test data: a collection of artifacts with \emph{a priori} known properties, such as the types and locations of vulnerabilities, which allows us to compare the results of an automated analysis against this ground truth.
Generated test data has the appeal of being comparatively cheap to obtain in large quantities and with controlled properties.
At the same time, there is a danger of missing essential features of security issues `in the wild'.

Real-world datasets are therefore attractive as a means of evaluating performance on cases that really occurred and may thus seem more realistic.
Finding such instances, however, is like looking for a needle in a haystack: in large real-world collections, the vast majority of artifacts are typically negative samples that do not exhibit the desired property.
Thus, unless there is already an established method for identifying positive instances, we depend on cases individually detected over time.
In this paper, we search one collection of such cases, the CVE database~\cite{CVE}, for all issues related to smart contracts.

\subsection*{Smart contracts}

An Ethereum-based blockchain is a decentralized network in which each node maintains the state of the blockchain~\cite{Wood}.
The state maps numeric addresses to accounts that include an Ether balance and, optionally, persistent data storage and a sequence of instructions known as a \emph{blockchain program} or \emph{smart contract}.
Whenever an account with contract code is called, its instructions are executed by the Ethereum Virtual Machine (EVM) and may, among other things, modify persistent storage and perform calls to other accounts.
Most smart contracts are written in the programming language Solidity, which compiles to EVM bytecode~\cite{Solidity}.
The source code specifies the smart contract at a higher level of abstraction and is usually essential for understanding its functionality.
Blockchain explorers such as Etherscan allow developers to upload the source code~\cite{EtherscanVerification}.
The web service then verifies that it compiles to the bytecode deployed on the blockchain and displays it alongside the account information.

Like any software, smart contracts are prone to errors.
Some bugs can be attributed to versioned products, such as libraries, that are maintained by an identifiable entity and for which a fixed version may be released.
Often, however, vulnerabilities are reported for source code uploaded to and displayed by a blockchain explorer for a specific deployed contract.
In such a case, the stakeholders (programmer, deployer, and users) are usually unknown.
Moreover, the immutability of deployed smart contracts is essential for applications such as tokens and may preclude fixing a bug by simply deploying an updated version at the same address.
Hence, vulnerability reports for individual deployments may be of limited value if there is no way to communicate them confidentially to a maintainer, if publication would facilitate attacks, and if no direct fix is available.

\subsection*{Common Vulnerabilities and Exposures (CVE)}

The \emph{Common Vulnerabilities and Exposures} program (CVE) aims at identifying, defining, and cataloging publicly disclosed cybersecurity vulnerabilities~\cite{cve_overview}.
Vulnerability reports are submitted to one of 500+ partner organizations called \emph{CVE Numbering Authorities} (CNAs), which check the formal consistency of the reports and assign unique CVE identifiers.
Records in the CVE database link these identifiers to short descriptions of the vulnerabilities.
A focus is on providing CVE identifiers quickly so that they can be used to correlate vulnerability data across tools, databases, and people.

In 2016, the DAO hack prompted the CVE board to briefly discuss whether smart contract vulnerabilities qualify for the CVE program, without reaching a conclusion~\cite{msg00022}.
Among the questions raised was whether smart contracts are customer-controlled software or rather online services, the latter being out of scope under the CNA rules at the time.
Another question was whether vulnerabilities in blockchain programs considered as legal contracts should be judged by the spirit or by the letter of the coded `law'.

The topic lay dormant until 2018, when the first CVE identifiers for actual smart contract vulnerabilities were requested.
In one case, a company mass-filed 401 reports, apparently employing a script to filter contracts for vulnerable code snippets.
Mitre, the CNA in charge, took a pragmatic approach and assigned about 500 identifiers.
The CVE board subsequently took up the discussion on its mailing list~\cite{msg00004} and in a board meeting~\cite{msg00026}.
Despite arguments that the filed vulnerabilities did not meet several criteria, the discussion concluded, without a formal vote, in favor of accepting such reports.
After the initial spike in 2018, the number of smart contract related submissions never exceeded 11 in any subsequent year.
We see reports about versioned libraries and projects maintained by companies, alongside only a handful of vulnerabilities reported for individual deployments.

\subsection*{Dataset}

The dataset \emph{CVE-Smart-Contracts} collects all CVE records up to 24 July 2026 (commit \stamp{9fdf581} of the CVE v5 repository~\cite{CVE}) that address a smart contract vulnerability.
For CVE records referring to a deployed contract without maintainer, the dataset provides the deployment address, the source code, the deployed bytecode, vulnerability labels according to three taxonomies, and information about the vulnerability location.
For projects (mostly libraries) maintained in conventional repositories, the dataset provides a manual snapshot of the vulnerable code and generated vulnerability labels.
The labels and locations characterize the reported claims; they do not independently establish vulnerability or exploitability.

Scripted pipelines make the collection, validation, and integration steps reproducible.
Crucially, a scoring algorithm evaluates the correspondence between CVE records and the artifacts they purportedly refer to.
Based on external documents and data retrieved from the blockchain (both archived in the dataset for reproducibility), scripts compute scores for two claims: (1) the CVE report pertains to the retained source, and (2) the named contract is deployed at the indexed address.
For cases with low scores, the correspondence is verified manually and the decision is recorded, with a justification, in the dataset.
In the end, we can establish for each CVE record whether its vulnerability claim can be reliably linked to the retained artifacts.
Typical causes of non-matches are erroneous or underspecified CVE records.

The complete curated dataset contains 568 CVE records, of which 541 refer to contracts deployed on Ethereum, 1 to a contract on the BNB Smart Chain, and 26 to libraries and packages.
For the 542 records referring to deployed contracts, correspondence validation yields 497 matching and 45 non-matching records.
Six of the matching records are marked as \emph{refuted} for reasons like the vulnerable function being commented out, rendering the CVE record a false positive.
Therefore, the top-level catalog \path{cve.json} lists 491 correspondence-matched, non-refuted contracts with vulnerability labels and locations.

\subsection*{Related work}

Previous collections of smart contracts derived from CVE records were primarily compiled for program analysis and repair.
The VeriSmart benchmark~\cite{VeriSmart, VeriSmartGit} contains 487 contracts associated with CVE records, whereas the SmartFix dataset~\cite{SmartTest, SmartFix, SmartFixGit} uses a subset of 200 VeriSmart contracts selected with a focus on program repair.
All CVE records from both datasets are fully included in the present dataset.
A file-level comparison with VeriSmart found 485 of 487 artifacts to be substantially the same (disregarding line endings, white space, pragma and modifier corrections, or bundles). Two cases select different artifacts, which are set to invalid in VeriSmart, while they pass correspondence validation (see below) in the present dataset.
The SmartFix dataset retains the VeriSmart source files while adding project-internal metadata.

The present dataset goes beyond these earlier corpora in two respects.
First, it is more comprehensive, as it identifies all CVE reports related to smart contracts through mid-2026.
Second, it substantially enriches the data: it provides the affected CVE records, collects and archives the external documents necessary to interpret and locate the reported vulnerability, adds the affected source code and runtime bytecode, and employs scripted pipelines for tasks such as data validation, assignment of vulnerability labels in three taxonomies, and function-level vulnerability localization.

\subsection*{Use cases}

The dataset is intended to support:
(1) studies of vulnerabilities reported in deployed or deployable smart contracts;
(2) the construction and auditing of analysis benchmarks;
(3) vulnerability classification and retrieval;
(4) research on function- and statement-level localization;
(5) the evaluation of static, dynamic, language-model-based, and repair systems; and
(6) studies of the quality and evolution of public vulnerability metadata.

\section*{Methods}

\begin{figure}[H]
\centering
\resizebox{0.94\linewidth}{!}{%
\begin{tikzpicture}[
  font=\small,
  data/.style={draw, rounded corners=2pt, fill=blue!8, align=center,
    minimum height=9mm, text width=25mm},
  action/.style={draw, rounded corners=5pt, fill=orange!15, align=center,
    minimum height=9mm, text width=25mm, font=\small\bfseries},
  decision/.style={draw, dashed, rounded corners=2pt, fill=black!4,
    align=center, minimum height=8mm, text width=25mm},
  catalog/.style={draw, rounded corners=2pt, fill=green!13, align=center,
    minimum height=10mm, text width=25mm, font=\small\bfseries},
  primary/.style={-{Latex[length=2mm]}, thick, rounded corners=4pt},
  dependency/.style={-{Latex[length=2mm]}, thin, draw=black!65,
    rounded corners=4pt},
  reviewed/.style={-{Latex[length=2mm]}, semithick, dashed}
]
\node[catalog] (cve)      at ( 0.0, 5.4) {CVE records};
\node[catalog] (code)     at (-6.0,-1.0) {Code artifacts};
\node[catalog] (match)    at ( 0.0,-1.0) {\texttt{match}/\texttt{non\_match}};
\node[catalog] (labels)   at ( 6.0,-1.0) {Labels};
\node[catalog] (locations) at ( 3.0,-1.0) {Locations};
\node[data] (previous) at (-5.0, 4.4) {Prior datasets};
\node[data] (taxonomy) at ( 4.8, 4.4) {Taxonomies};
\node[data]   (evidence) at (-3.0,-1.0) {Evidence};
\node[action] (acquire)  at (-6.0, 2.0) {Acquire};
\node[action] (validate) at (-2.0, 2.0) {Validate};
\node[action] (localize) at ( 2.0, 2.0) {Localize};
\node[action] (label)    at ( 6.0, 2.0) {Assign labels};
\draw[primary] (previous) -- (acquire);
\draw[primary] (cve) -- (acquire);
\draw[primary] (acquire) -- (code);
\draw[primary] (acquire) -- (evidence);
\draw[primary] (cve) -- (validate);
\draw[primary] (code) -- (validate);
\draw[primary] (evidence) -- (validate);
\draw[primary] (validate) -- (match);
\draw[primary] (cve) -- (label);
\draw[primary] (evidence) -- (label);
\draw[primary] (taxonomy) -- (label);
\draw[primary] (label) -- (labels);
\draw[primary] (cve) -- (localize);
\draw[primary] (match) -- (localize);
\draw[primary] (code) -- (localize);
\draw[primary] (evidence) -- (localize);
\draw[primary] (localize) -- (locations);

\node[catalog, text width=14mm, minimum height=6mm] at (-2.0,-3.0) {catalog};
\node[data, text width=14mm, minimum height=6mm] at ( 0.0,-3.0) {data};
\node[action, text width=15mm, minimum height=6mm] at ( 2.0,-3.0) {action};
\end{tikzpicture}%
}
\caption{Workflow of dataset construction. Green items denote components for the final catalog \texttt{cve.json}, while blue items represent additional data. Actions (orange) drive the workflow and pass on data to subsequent steps. Catalog construction has been omitted as a separate action, as it only combines the green components.}
\label{fig:data-lineage}
\end{figure}
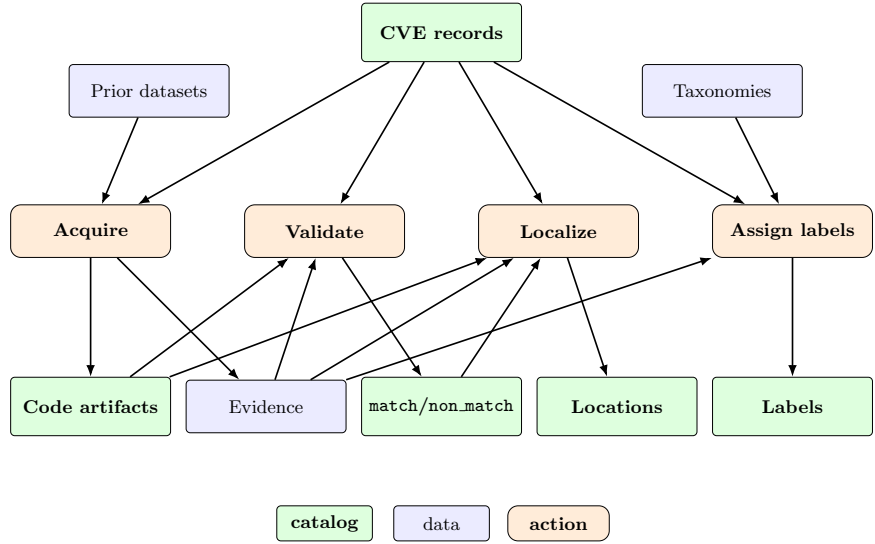

Dataset construction consists of five phases (Fig.~\ref{fig:data-lineage}).
\emph{Data acquisition} collects and organizes the core data.
A central task is to associate CVE records with the artifacts they refer to. 
\emph{Correspondence validation} corroborates this association by analyzing the external documents referenced by the CVE records and obtaining further evidence from the blockchain.
\emph{Localization} records where the retained evidence places each claim within the source code by specifying the affected functions.
\emph{Labeling} maps the verbal vulnerability descriptions to categories in structured taxonomies.
\emph{Catalog construction} joins the information obtained in the preceding phases and makes the results available as a succinct JSON file.

\subsection*{Data acquisition}

We start from a copy of the CVE database~\cite{CVE}, commit \stamp{9fdf581} dated 24 July 2026.
We consider records in scope if they pertain to code for an EVM-compatible contract or library, and exclude records pertaining to client software, development tools such as compilers, off-chain applications, or protocol-level issues without an affected contract artifact.

The script \path{filter_ethereum_smart_contract_cves.py} selects candidate records from the CVE database using text and reference cues such as \emph{smart contract}, \emph{Ethereum}, \emph{Solidity}, names of token standards, and explorer links.
The resulting selection includes the records represented in the VeriSmart benchmark at commit \stamp{8cbb2db}, dated 17 August 2022, and the SmartFix artifact at commit \stamp{d2db657}, dated 21 August 2023~\cite{VeriSmart, SmartTest, SmartFix}.
We manually assessed the additional candidates and excluded those out of scope, so that the filter script now yields exactly the 568 CVE records retained in the dataset.
We cannot guarantee completeness, however, as a record within scope may fail to include any of the cues.

Where possible, we associate the CVE record with a deployment address, either given in the record itself or mentioned in an external document.
To simplify the collection of deployment addresses, we reuse data from prior datasets.
With the address as an argument, the script \path{fetch_etherscan_artifacts.py} accesses the Etherscan API V2 endpoint~\cite{Etherscan} using the \path{contract/getsourcecode}, \path{proxy/eth_getCode}, and \path{proxy/eth_call} actions, and downloads source code, compilation metadata, runtime bytecode, and the return values of certain view functions of token contracts.
Maintained and versioned projects, in particular libraries, are treated differently, as they refer to a specific version in a repository rather than a deployment.
Here, we retain a copy of the version that is flagged as vulnerable.

The curated file \path{data/index.csv} relates the CVE identifier to the focal artifact, contract name, chain, source address, deployment address, refuted status, and explanatory notes.
Repository versions are listed in \path{REPRODUCIBILITY.md}.
The dataset includes third-party source and evidence, when necessary to support artifact identity, validation, and reproducibility. Original notices and available license information are retained.
\path{THIRD_PARTY_NOTICES.md} documents the provenance and licensing boundary; the dataset and original project code are licensed separately.

\subsection*{Correspondence validation}

Validation asks whether retained evidence supports the dataset's mapping from a CVE description to a source artifact and, where given,
a deployment.
A \textit{match} does not independently confirm that the code is vulnerable; a \textit{non-match} rejects the indexed correspondence rather
than the CVE claim itself.

More specifically, the file \path{data/index.csv} contains 542 rows that refer to a deployed contract.
For each of these rows, the fields \emph{CVE identifier}, \emph{contract name}, \emph{chain}, \emph{source address} and \emph{deployment address} encode two claims.
\begin{itemize}
\item The vulnerability description in the CVE record refers to the source code downloaded from Etherscan for the indicated source address.
\item The named contract with the purported vulnerable code has been deployed at the indicated deployment address.
\end{itemize}
The distinction between source and deployment address is relevant because the source code specified by the address in the CVE record may contain multiple contracts, of which the contract identified by the claim may be deployed at a different address or not at all.

The correspondence rules classify 450 records automatically as matches, since the chain address can be located in the CVE record or in supporting documents, the names of functions, contracts, and tokens mentioned in the CVE record match the source code and the data queried from Etherscan, and a supporting document lists a code snippet that occurs verbatim in the downloaded source code.
The rules also classify 39 records automatically as non-matches, since the retained source contradicts the indexed contract or the named function.
The remaining 53 records require a reviewed decision as the available observations are incomplete, contradictory, or not covered by an automatic rule.
These manual decisions classify 47 records as matches and six as non-matches. The final validation data therefore contain 497 matches and 45 non-matches.

\subsection*{Vulnerability labeling}
The vulnerability labels map each reported vulnerability claim to identifiers in three taxonomies. 
They characterize the claim represented by the retained evidence and do not independently establish that a weakness exists or is exploitable in a particular artifact or deployment.

\subsubsection*{Taxonomies employed}
We employ the smart contract vulnerability taxonomy of Iuliano and Di Nucci~\cite{taxonomy}, the Smart Contract Weakness Classification Registry (SWC)~\cite{SWCregistry}, and the Common Weakness Enumeration (CWE)~\cite{CWE}. 
The first provides a recent, comprehensive, and well-researched hierarchical taxonomy developed for smart contract weaknesses. 
The SWC taxonomy is no longer maintained, but still popular and widely used in the smart contracts community.
CWE describes itself as a ``community-developed list of common software and hardware weaknesses'' and provides the general software-weakness identifiers used by the CVE program.

The three label assignments are carried out independently; the SWC mappings mentioned in the Iuliano and Di Nucci taxonomy are neither explicitly nor implicitly used for the SWC assignments. 
The frozen inputs use Table 10 of the Iuliano and Di Nucci taxonomy, SWC Registry commit \stamp{1b62270} dated 6 August
2024, and CWE catalog 4.20 dated 30 April 2026. 
Their machine-readable representations and provenance are retained under \path{labeling/taxonomies/}.

\subsubsection*{Labeling process}
Labeling consumes the retained CVE records in \path{data/cves/} and the provenance-preserving evidence corpus \path{validation/generated/cve_texts.jsonl}. 
The label generator reads a set of 13 declarative rules from \path{labeling/rules/vulnerability_rules.json}.
Each rule specifies constraints that determine its applicability as well as primary and secondary assignments for the Iuliano and Di Nucci, CWE, and SWC taxonomies.

The rules are applied to eligible text ordered into six tiers: CVE descriptions, CVE titles, manually curated evidence, directly referenced documents, text derived from reference metadata or optical character recognition, and source artifacts. 
The strongest tier containing a rule match determines the suggestion, while matches in weaker tiers are retained only as diagnostics.

Each constraint is a list of regular expressions applied case-insensitively.
There are three types: an all-constraint is satisfied if all expressions in the list match, an any-constraint requires at least one expression to match, and a none-constraint succeeds only when none of the expressions in the list match.

Reference evidence is eligible only when its provenance records a sufficiently strong relationship to the CVE. Boilerplate, bare reference URLs, and unrelated evidence are excluded. 
Source code can contribute only in the final evidence tier and only through a rule that explicitly permits code evidence.

A single applicable rule at the deciding tier produces an automatic suggestion with assignments for all three taxonomies.
Multiple applicable rules at that tier are treated as ambiguous rather than ranked, and absence of an applicable rule produces an insufficient-evidence result.
The generator also preserves CWE assertions supplied by the CVE record and reports whether they agree with the suggestion, but it does not automatically adopt them as the final CWE label.

Label generation produces 524 automatic suggestions and 44 results with insufficient evidence. 
Reviewed decisions under \path{labeling/decisions/} cover the 44 abstentions and replace 12 automatic suggestions. 
Each decision specifies all three taxonomy assignments and a justification. 
A reviewed record can remain \textit{not classifiable}, in which case all three primary assignments are null and the secondary lists are empty.

The script \path{labeling/scripts/build_labels.py} validates every identifier against the frozen taxonomy representations, gives reviewed decisions precedence over suggested labels, and writes \path{labeling/labels.json}. 
The released file contains 512 rule-based classifications, 55 manually classified records, and one manually reviewed record retained as \textit{not classifiable}.
It records the applicable rule and evidence tier or the path to the reviewed decision, so the origin of every resolved label is machine-readable.

\subsection*{Vulnerability localization}
Localization attributes each non-refuted matching contract claim to one or more function-like units in the retained verified source.
It reflects where the CVE record and its supporting evidence place the claim; it is not an independent finding that the indicated code is vulnerable.

The script \path{localization/scripts/extract_source_units.py} selects non-refuted contract rows classified as \textit{match} by correspondence validation, picks the corresponding JSON file in \path{data/etherscan} with contract source and compilation metadata, and uses an appropriate version of the Solidity compiler\footnote{%
  Appropriate here means the version indicated by compilation metadata, except for 0.3.6 replacing the unavailable version 0.3.4.%
} to generate the abstract syntax tree (AST).
It normalizes legacy and compact ASTs into an inventory of source units, inheritance relations, functions, constructors, fallback and receive functions, modifiers, visibility, modifier applications, and statically resolvable direct calls, with corresponding line ranges.

Candidate generation combines this inventory with the provenance-preserving evidence corpus.
It extracts contract, function, and modifier names occurring in the inventory and compares evidence code blocks with individual source units using token-preserving matching.
Direct call and modifier edges are traversed backwards to identify externally callable entry points.
Candidate scores use the strongest occurrence of each signal type: exact code matches rank above explicit function names, and contract-name evidence establishes scope but is not sufficient on its own to select a function.
The resulting confidence values express evidential support for a location, not the probability that the vulnerability claim is true.

Generated candidates are accepted without a reviewed decision only for high-confidence leaf functions declared by or inherited into the indexed contract, or for an uncontested medium-confidence function declared by that contract.
This conservative automatic procedure supplies locations for 415 records.
Ambiguous candidates, call-chain locations, contract conflicts, analysis failures, and cases without a candidate require review; locations for 76 records are supported by manual decisions.
Each such decision records a summary and supporting evidence and identifies the source path, contract, canonical function signature, line range, optional statement annotations, and entry-point paths.
A reviewed decision may instead record an unresolved outcome and its reason rather than force a source location; unresolved decisions remain in the review provenance and are omitted from the released mapping.
All 491 records selected for the present catalog have a released location record.


\subsection*{Dataset integration and versioning}
The integrated catalog \path{cve.json} is generated rather than manually edited.
Selection begins with contract rows in \path{data/index.csv} that are not refuted and are classified as \textit{match} in \path{validation/validation.csv}.
Of the 497 matches, six are refuted, leaving 491 selected records.
The builder requires a resolved record in \path{labeling/labels.json} and a released record in \path{localization/locations.json} for every selected CVE; missing component records stop the build.

For each selected CVE, the builder checks that the retained CVE record, affected source artifact, and verified source from Etherscan exist.
When a deployment address is available, it also requires the corresponding runtime bytecode.
It then combines repository-relative artifact paths, source and deployment identities, the three resolved taxonomy assignments, and the location data into a CVE-keyed record.

After the catalog is built, \path{scripts/build_metadata.py} derives \path{metadata.json} from the index and component outputs.
It records the dataset and schema versions, release and CVE snapshot dates, calculated coverage counts, taxonomy provenance, selection rules, and schema locations.
The release version and date are maintained explicitly, while coverage counts are recomputed rather than copied from prose.
The DOI-bearing archive is the citable frozen release; the GitHub repository is the maintained development version, in which later evidence or annotation corrections may appear.

\subsection*{Reproducible build environment}
The normal build consumes retained acquisition inputs and versioned OCR text; it does not refresh CVE, explorer, blockchain, reference, or OCR material from the network.
Network acquisition and platform-dependent OCR refreshes are separate maintainer operations because their results can change over time.

Python dependencies are locked in \path{requirements.txt}, and the \texttt{ethutils} dependency is pinned to commit \stamp{b2c349a}.
The reference environment uses Python 3.14.3 on macOS Darwin 25.5.0 arm64.
Exact Solidity compiler releases reported by the retained verified-source responses are selected with \texttt{solc-select}; missing required releases are installed, and a failure to install, select, or run a compiler stops the pipeline.
Releases before Solidity 0.8.5 require Rosetta on Apple Silicon, and the Solidity 0.3.6 substitution required for the oldest source is unavailable through the supported compiler inventory on Linux.

The deterministic build order is correspondence validation, labeling, localization analysis and generation, integrated catalog construction, and metadata generation.
Machine-readable schemas are checked separately or as part of the release verification script.
The verification script executes these stages from a clean tracked worktree and rejects any difference between regenerated and committed outputs.

\section*{Data Records}

The CVE-Smart-Contracts dataset is maintained on GitHub~\cite{CVE-Smart-Contracts} and deposited as a versioned archive on Zenodo~\cite{zenodo}.
The archived dataset version described here is \stamp{2026-08-14}.
Except for the third-party materials identified below, the dataset and original research content are licensed under Creative Commons Attribution 4.0; original repository software is licensed under the MIT License.
Table~\ref{tab:data-files} lists the main components of the dataset and their role.

\begin{table}[ht]
\centering
\small
\begin{tabularx}\linewidth{p{8.5em}p{3em}X}
\toprule
Path & Format & Population and role \\
\midrule
\path{cve.json} & JSON & Integrated catalog of 491 matched, non-refuted, labeled, and localized contract records \\
\path{metadata.json} & JSON & Release identity, CVE snapshot, calculated coverage, taxonomy provenance, selection rules, and schema paths \\
\path{data/} &  & Retained CVE, source, explorer-response, and runtime inputs. \\
\path{data/index.csv} & CSV & Complete curated population of 568 CVE-to-artifact mappings \\
\path{validation/}\newline\hspace*{1em}\path{validation.csv} & CSV & Correspondence checks and final classification for all 542 contract rows. \\
\path{validation/}\newline\hspace*{1em}\path{evidence/} &  & Retained reference, call-result, manual, and OCR evidence. \\
\path{labeling/}\newline\hspace*{1em}\path{labels.json} & JSON & Resolved three-taxonomy assignments and provenance for all 568 indexed CVEs. \\
\path{localization/}\newline\hspace*{1em}\path{locations.json} & JSON & Detailed released source locations for the 491 catalog records. \\
\path{*/decisions/} & JSON & Reviewed validation, labeling, and localization decisions. \\
\path{*/generated/} & JSON,\newline JSONL & Reproducible intermediate analyses and candidates. \\
\bottomrule
\end{tabularx}
\caption{Principal released files and their roles. Detailed field
specifications and JSON Schemas are retained beside the corresponding files.}
\label{tab:data-files}
\end{table}

\subsection*{Curated index and retained artifacts}

The file \path{data/index.csv} is the broadest machine-readable population.
It contains 568 CVE records, comprising 542 contract records and 26 project (library) records.
It retains refuted claims and rejected artifact correspondences so that selection decisions remain auditable.
Each row associates one canonical \code{cve_id} with an affected artifact.
The nine fields are specified in Table~\ref{tab:index-fields}.

\begin{table}[ht]
\centering
\small
\begin{tabularx}\linewidth{l>{\raggedright\arraybackslash}p{0.2\linewidth}X}
  \toprule
  Field & Format & Meaning \\
  \midrule
  \code{cve_id} & \code{CVE-20}$yy$\code{-}$n\cdots n$ & the canonical CVE identifier \\
  \code{refuted} & \code{t} or \code{f} & true when review determined that the reported vulnerability claim does not hold; false otherwise \\
  \code{kind} & \code{contract} or \code{library} & type of artifact, deployed contract or versioned project (library) \\
  \code{artifact} & rel.path & affected source file or directory (in dataset) \\
  \code{contractname} &    & name of affected Solidity contract \\
  \code{chain} & \code{ethereum} or \code{bsc} & chain for \code{src_addr} and \code{addr} \\
  \code{src_addr} & \code{0x}+hex digits & address for which verified source was obtained \\
  \code{addr} & \code{0x}+hex digits &  the claimed affected deployment address \\
  \code{note} &&  an optional row-specific comment \\
\bottomrule
\end{tabularx}
\caption{Structure of the curated index \texttt{data/index.csv}}
\label{tab:index-fields}
\end{table}

For every CVE identifier, the folder \path{data/cves/} contains the original CVE v5 record, whereas the folders \path{data/contract_sources/} and \path{data/library_sources/} contain a file or directory with the Solidity sources, as specified by the \code{artifact} column of \path{index.csv}.
For every address mentioned in \path{index.csv}, the folders \path{data/etherscan/} and \path{data/runtime/} contain the corresponding source code and runtime bytecode, respectively.
For addresses, the source code is stored in form of a JSON file as obtained from Etherscan that includes compilation information as well.

\subsection*{Integrated catalog}

The integrated catalog \path{cve.json} contains the 491 non-refuted contract records whose artifact correspondence is classified as a match and for which resolved labels and released location data are available.
It is the recommended entry point when a compact set of correspondence-matched, labeled, and localized contract cases is required.
Table~\ref{tab:cve-fields} describes its structure.
The formal specification is available as \path{scripts/cve.schema.json}.

\begin{table}[ht]
\centering
\small
\begin{tabularx}\linewidth{lX}
  \toprule
  \code{.artifacts} & information on artifacts \\
  \quad\code{.cve} & path to CVE record \\
  \quad\code{.source} & chain, address, contractname, path to source code \\
  \quad\code{.runtime} & chain, address, path to runtime bytecode \\
  \code{.labels} & primary and secondary labels \\
  \quad\code{.IulianoDiNucci2026}& \dots\ in Iuliano-DiNucci taxonomy \\
  \quad\code{.CWE} & \dots\ in CVE taxonomy \\
  \quad\code{.SWC} & \dots\ in SWC taxonomy \\
  \code{.localization} & information on vulnerability location\\
  \quad\code{.summary} & prose description\\
  \quad\code{.locations[]} & list of locations \\
  \quad\quad\code{.source_path} & source file \\
  \quad\quad\code{.contract} & contract name \\
  \quad\quad\code{.function} & vulnerable function (internal or external)\\
  \quad\quad\code{.entry_points[]} & external entry points that use the vulnerable function \\
\bottomrule
\end{tabularx}
\caption{Structure of the object summarizing the information for each CVE entry in the catalog \texttt{cve.json}, accessible via \texttt{.records[\textit{cve-identifier}]}.}
\label{tab:cve-fields}
\end{table}

The \code{labels} object contains parallel assignments for the Iuliano-DiNucci, CWE, and SWC taxonomies.
Each assignment has a primary label and a possibly empty list of secondary labels.
A primary value is null if the corresponding taxonomy has no supported assignment.
The provenance of the classification~-- rule vs.\ manual decision~-- is available in the labels file \path{labeling/labels.json} and not duplicated here.

The \code{localization} object contains a summary and one or more locations.
Each location identifies an embedded source path, declaring contract, function or modifier, inclusive function line range, and public or external entry-point
paths through which the location is exposed.

\subsection*{Dataset metadata}

The generated \path{metadata.json} file records the dataset name, version,
release date, and description; the CVE Record Format 5 source repository,
commit, commit date, and latest retained-record update; and calculated counts
for index kinds, chains, refuted records, validation classes, labels, location
records, and catalog records. It also provides the frozen taxonomy
sources, textual selection rules, and repository-relative paths to all JSON
Schemas. Coverage values are derived from current component files by
\path{scripts/build_metadata.py}; they are not maintained independently in
documentation.

\subsection*{Correspondence validation data}

The file \path{validation/validation.csv} contains one row for each of the 542 contract records.
Each row starts with the CVE identifier as its key, followed by 20 Boolean flags, the classification result (\code{match} or \code{non_match}), the classification reason (a rule name or \code{manual}), and an optional note with the rationale of manual decision.
The flags are the results of elementary checks that form the basis for classification.
As an example, the flag \code{token_name} is true, if the token name as queried from the blockchain for the source or deployment address can be found in the evidence material, false if it is not found, and null if the queried contract does not implement the \code{name()} function.
For a detailed description of the flags and their use, see \path{validation/validation.md}.

Detailed observations and compiled interfaces are retained under \path{validation/generated/}, while durable acquired evidence and manual captures are under \path{validation/evidence/}.
Reviewed overrides conform to \path{validation/decision.schema.json} and are stored under \path{validation/decisions/}.

\subsection*{Vulnerability locations}

The file \path{localization/locations.json} is a CVE-keyed mapping for the 491 catalog records.
It is the source of the function-level locations in the \code{localization} objects of the catalog \path{cves.json}.
Additionally, \path{locations.json} offers statement-level locations.
However, as only three CVE records contain sufficient data to identify the line of the vulnerability, this field is not exported to the top-level catalog.

The directory \path{localization/decisions/} retains reviewed decisions and their supporting evidence.
Compiler-derived source units, evidence indicators, ranked candidates, candidate decisions, and warnings are retained under \path{localization/generated/}.
The released locations conform to \path{localization/locations.schema.json}, and reviewed decisions conform to \path{localization/decision.schema.json}.

\subsection*{Vulnerability labels}

The file \path{labeling/labels.json} is a CVE-keyed mapping for all 568 indexed CVE records, with its format specified in \path{labeling/labels.schema.json}.
It is the source of the \code{labels} objects in the catalog \path{cves.json}, but also provides labels for the other CVE records in \path{index.csv}.
Additionally, \path{labels.json} contains information about the labeling process, like which rule was applied on which evidence tier or where a manual decision is recorded.
All but one CVE records contain sufficient information to derive vulnerability labels from the record and supporting evidence material, getting marked as \code{classified}.
Only one CVE record remains \code{not_classifiable}, since the CVE description is inconclusive, the referenced documents cannot be retrieved, and the associated artifact does not match.

Upstream CNA and ADP CWE assertions and their comparison with generated suggestions remain in \path{labeling/generated/label_suggestions.json}; they are not silently substituted for the resolved CWE assignment.
The automatic classification uses declarative rules that are stored in \path{labeling/rules/vulnerability_rules.json}; for a discussion of how rules are applied see the methods section.
The frozen taxonomies and their provenance can be found as JSON files under \path{labeling/taxonomies/}.
The 56 reviewed decisions with a justification are retained in the folder \path{labeling/decisions/}, with the format specified in \path{labeling/label_decision.schema.json}.

\subsection*{Evidence, decisions, and generated intermediates}

The repository distinguishes retained inputs, reviewed decisions, and reproducible intermediates.
The \path{validation/evidence/} tree contains archived references and their provenance, normalized documents, retained images and versioned OCR results, blockchain call results, and curated sources.
Decision directories contain the current human judgments that replace or supplement automatic results.
Generated directories contain compiler interfaces, runtime selectors, evidence corpora, match observations, classification suggestions, AST inventories, and localization candidates that can be rebuilt from the retained inputs.

This distinction is important for reuse: generated files explain how a result was obtained, decision files explain reviewed exceptions, and retained evidence permits the interpretation to be audited without depending on a live external
page.
Network refreshes are intentionally separate operations and may produce different retained inputs at a later date.

\section*{Data Overview}
\begin{figure}[H]
\centering
\begin{tikzpicture}[
  font=\small,
  population/.style={draw, rounded corners=2pt, fill=blue!8,
    align=center, minimum height=9mm, text width=29mm},
  selected/.style={draw, rounded corners=2pt, fill=green!12,
    align=center, minimum height=9mm, text width=29mm},
  excluded/.style={draw, dashed, rounded corners=2pt, fill=black!4,
    align=center, minimum height=9mm, text width=27mm},
  flow/.style={-stealth, thick, gray, shorten <=3pt, shorten >=3pt}
]
\node[selected] (all) {568 CVE records\\\texttt{data/index.csv}};
\node[population,below left=1 and -1 of all] (contracts) {542 contract records\vphantom{y}};
\node[population,below right=1 and -1 of all] (libraries) {26 library records};

\node[population, below left=1 and 0 of contracts] (matches) {497 correspondence\\ matches};
\node[population, below right=1 and 0 of contracts] (nonmatches) {45 correspondence non-matches};

\node[selected, text width=23mm, below left=1 and -1.3 of matches] (catalog) {491 non-refuted matches\\\texttt{cves.json}};
\node[population, text width=23mm, below right=1 and -1.3 of matches] (matchrefuted) {6 refuted matches};
\node[population, text width=23mm, below left=1 and -1.3 of nonmatches] (nonrefuted) {43 non-refuted non-matches};
\node[population, text width=23mm, below right=1 and -1.3 of nonmatches] (nonmatchrefuted) {2 refuted non-matches};

\draw[flow] (all) -- (contracts);
\draw[flow] (all) -- (libraries);
\draw[flow] (contracts) -- (matches);
\draw[flow] (contracts) -- (nonmatches);
\draw[flow] (matches) -- (catalog);
\draw[flow] (matches) -- (matchrefuted);
\draw[flow] (nonmatches) -- (nonrefuted);
\draw[flow] (nonmatches) -- (nonmatchrefuted);
\end{tikzpicture}
\caption{Population flow from the curated index to the integrated
catalog.}
\label{fig:population-flow}
\end{figure}
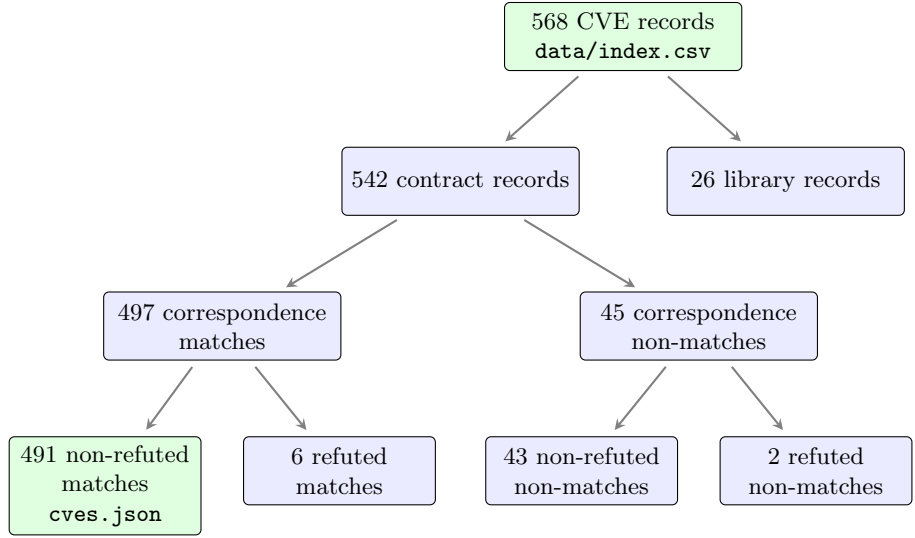

The population flow in Fig.~\ref{fig:population-flow} shows how the complete curated index relates to the integrated catalog.
Table~\ref{tab:vuls-per-year} gives an overview of all CVE entries and their vulnerabilities by year and reporter.
Table~\ref{tab:cves-vuls} lists the vulnerabilities retained in \path{cves.json} ranked by their frequency.

\begin{table}[htp]
  \newcommand\ROW[7]{#1 & #2 & #3 & #7 & #4 & #5 & #6 \\}
  \centering\footnotesize
  \setlength\tabcolsep{3pt}
  \begin{tabularx}\linewidth{rlrrrrX}
    \toprule
    \ROW{Year}{Reporter}{Cnt}{Nm}{Ref}{Top vulnerabilities}{Prj}
    \midrule
\ROW{2018}{BlockChainsSec}{401}{42}{2}{6A1 Int overflow (401)}{}
\ROW{}{VenusADLab}{32}{}{}{6A1 Int overflow (32)}{}
\ROW{}{PeckShield}{22}{}{}{6A1 Int overflow (10), 5F Overpowered Role (7)}{}
\ROW{}{dwfault}{10}{}{}{6A1 Int overflow (10)}{}
\ROW{}{safecomet}{8}{}{4}{6A1 Int overflow (8)}{}
\ROW{}{jonghyk.song}{7}{}{}{2C Bad random number generation (4)}{}
\ROW{}{hellowuzekai}{6}{}{}{6A1 Int overflow (4)}{}
\ROW{}{TEAM-C4B}{6}{}{}{2C Bad random number generation (5)}{}
\ROW{}{n0pn0pn0p}{5}{}{}{6A1 Int overflow (5)}{}
\ROW{}{SCResearcher}{5}{}{}{5B Acc.Cntr.Man. (3), 9C Err.~contr.~name (2)}{}
\ROW{}{rootclay}{2}{}{}{}{}
\ROW{}{Coinmonks}{2}{}{}{}{}
\ROW{}{others}{9}{}{}{6A1 Int overflow (4), 6A2 Int underflow (2)}{}
    \midrule
\ROW{2019}{smsecgroup}{3}{}{}{9C Err.~contr.~name (3)}{}
    \midrule
\ROW{2020}{hellowuzekai}{2}{}{}{}{}
\ROW{}{BlockSec}{1}{}{}{}{}
\ROW{}{ensdomains}{1}{}{}{}{1}
\ROW{}{others}{4}{}{}{}{}
    \midrule
\ROW{2021}{MRdoulestar}{4}{}{}{6A1 Int overflow (2)}{}
\ROW{}{OpenZeppelin}{4}{}{}{11E Privilege Escalation (2)}{4}
\ROW{}{BlockSec}{2}{}{}{}{}
    \midrule
\ROW{2022}{OpenZeppelin}{7}{}{}{10B Unhandled Exceptions in Library (2)}{7}
\ROW{}{others}{1}{}{}{}{}
    \midrule
\ROW{2023}{OpenZeppelin}{7}{}{}{7I7 Missing input validation (2)}{7}
\ROW{}{RikkaLzw}{3}{1}{}{}{}
\ROW{}{ensdomains}{1}{}{}{}{1}
    \midrule
\ROW{2024}{Wzy-source}{4}{1}{1}{}{}
\ROW{}{OpenZeppelin}{1}{}{}{}{1}
    \midrule
\ROW{2025}{OpenZeppelin}{1}{}{}{}{1}
\ROW{}{RikkaLzw}{1}{}{}{}{}
\ROW{}{others}{4}{}{}{}{3}
    \midrule
\ROW{2026}{ensdomains}{1}{}{}{}{1}
\ROW{}{others}{1}{1}{1}{}{}
    \midrule
\ROW{}{Total}{568}{45}{8}{}{26}
    \bottomrule
  \end{tabularx}
  \caption{Vulnerabilities by year and reporter. Reporters are
    identified by the Github or Medium user owning the supporting
    documents. \emph{Cnt} gives the total count, \emph{Prj} is the
    number of projects/libraries, \emph{Nm} the number of non-matches,
    \emph{Ref} the number of refuted claims, and \emph{Top
      vulnerabilities} lists vulnerabilities occurring multiple times,
    with the multiplicity in parentheses.}
  \label{tab:vuls-per-year}
\end{table}

\begin{table}[htp]
  \newcommand\ROW[5]{#1 & #2 & #3 & #4 & #5  \\}
  \centering\footnotesize
  \setlength\tabcolsep{3pt}
  \begin{tabular}{rl@{\ }lll}
    \toprule
    Cnt & \multicolumn{2}{l}{Iuliano-DeNucci-2026} & CWE & SWC \\
    \midrule
\ROW{431}{6A1}{Integer overflow}{190}{101}
\ROW{12}{5B}{Access Control management}{284, 862}{105}
\ROW{10}{2C}{Bad random number generation}{338}{120}
\ROW{8}{9C}{Erroneous constructor name}{665}{118}
\ROW{7}{5F}{Overpowered Role}{269}{}
\ROW{5}{6C1}{Incorrect Calculation}{682, 697}{101}
\ROW{4}{1A}{Call to the unknown}{749}{}
\ROW{3}{6A2}{Integer underflow}{191}{101}
\ROW{1}{11B}{Erroneous Accounting}{123}{}
\ROW{1}{11H}{Business Logic Flaw}{269}{}
\ROW{1}{1A1}{Reentrancy}{841}{107}
\ROW{1}{1D}{Vulnerable Delegatecall}{829}{112}
\ROW{1}{1H}{Send to zero address}{20}{}
\ROW{1}{3E}{DoS with failed call}{703}{113}
\ROW{1}{4I1}{DoS Costly Pattern and Loops}{400}{128}
\ROW{1}{5A1}{Authorization via transaction origin}{346}{115}
\ROW{1}{5B6}{Bypassable Modifier}{697}{}
\ROW{1}{5C1}{Missing protection against signature replay attack}{294}{121}
\ROW{1}{5C2}{Lack of proper signature verification}{863}{122}
\ROW{1}{7I}{Missing logic}{841}{}
\midrule
\ROW{491}{}{Total}{}{}
    \bottomrule
  \end{tabular}
  \caption{Frequency of the vulnerabilities in \texttt{cves.json}}
  \label{tab:cves-vuls}
\end{table}

\section*{Technical Validation}
Technical validation addresses five properties of the release: structural validity, consistency between data layers, evidential support for artifact correspondence, validity of taxonomy labels and source locations, and reproducibility of generated outputs.
These checks validate the dataset's representation of CVE claims and associated artifacts; they do not independently confirm that a reported vulnerability exists or is exploitable.

\begin{table}[ht]
\centering
\small
\begin{tabularx}\linewidth{lXX}
\toprule
Layer & Acceptance criterion & Failure behavior \\
\midrule
Structure
& Valid schemas, identifiers, field values, and repository-relative paths.
& Validation or build stops before publication of derived outputs.
\\
Cross-file integrity
& Equal CVE key sets and recomputed coverage and catalog selection.
& Build stops on a missing component, inconsistent key, or count mismatch.
\\
Artifact mapping
& An automatic rule or rationale-bearing reviewed decision resolves every contract record.
& Unresolved or contradictory evidence requires a reviewed decision; the result may be \texttt{non\_match}.
\\
Labels
& Valid taxonomy identifiers and one resolved label record per indexed CVE.
& Ambiguity or insufficient evidence requires a reviewed decision, which may retain \texttt{not\_classifiable}.
\\
Locations
& Valid source membership, ranges, signatures, entry points, and current input fingerprint.
& Unsafe automatic candidates require review; unresolved cases are excluded from the integrated catalog.
\\
Rebuild
& Regenerated tracked outputs equal the committed release.
& \path{verify_rebuild.sh} exits unsuccessfully on any tracked difference.
\\
\bottomrule
\end{tabularx}
\caption{Validation layers, acceptance criteria, and pipeline response to a failed criterion.}
\label{tab:validation-layers}
\end{table}

\subsection*{Structural and referential checks}

The integrity checks require the index to use its documented columns, unique CVE identifiers, enumerated record kinds and refutation values, normalized deployment addresses, and artifact names consistent with their CVE identifiers.
Every indexed artifact path must resolve within the repository, and every retained CVE record must be represented in the index.
Etherscan files are checked for required response fields, while runtime artifacts must contain normalized hexadecimal bytecode.

JSON records for the main catalog, metadata, labels, locations, and reviewed decisions are validated against the supplied JSON schemas.
The build scripts also require CVE-key consistency across the index, component records, validation results, labels, locations, and integrated catalog.
They recompute the catalog predicate from row-level data rather than accepting reported totals: a catalog entry must be a contract classified as a correspondence match and not marked as refuted.
The resulting key set contains 491 records, equal to the 497 correspondence matches after excluding the six matched records marked as refuted.

\subsection*{Source and deployment checks}
Correspondence validation combines CVE identity, retained source, compiler-derived interfaces, named-function evidence, verified-source metadata, deployment addresses, and selectors extracted from deployed runtime bytecode.
Source correspondence and deployment correspondence are represented separately, so the absence of a deployment address does not invalidate an otherwise supported source association.
The rule-based stage classified 450 records as matches and 39 as non-matches.
The remaining evidence was resolved in documented decisions, producing 47 reviewed matches and six reviewed non-matches; each reviewed decision retains its rationale and supporting evidence.
Together these checks yield 497 matches and 45 non-matches among the 542 contract records.

\subsection*{Label checks}
The label generator applies 13 ordered rules to the retained claim text and metadata and validates emitted identifiers and names against frozen versions of the Iuliano--Di Nucci, CWE, and SWC taxonomies.
It produced 524 automatic suggestions and 44 insufficient-evidence results.
Reviewed decisions take precedence over generated candidates and retain a justification; this includes 12 reviewed overrides of an automatic suggestion.
The released labels comprise 512 automatic classifications, 55 manually classified records, and one manually reviewed record marked \texttt{not\_classifiable}.
Each of the 568 indexed records therefore has a resolved label record, while null taxonomy values explicitly denote that no justified mapping was available rather than a negative finding.

\subsection*{Localization checks}
The localization pipeline validates source-unit membership, contract and function identity, compiler-derived source boundaries, inclusive line ranges, normalized signatures, statement anchors, and entry-point reachability.
Conservative automatic rules produced released locations for 415 records without a per-CVE reviewed decision; 76 further records have documented reviewed \texttt{localized} decisions.
Reviewed decisions take precedence over automatic candidates and retain their supporting evidence.
Input fingerprints prevent an annotation reviewed against an earlier artifact from being silently reused after the source or evidence changes.
All 491 catalog records have a released location record.

\subsection*{Reproducible build}
Python dependencies and the \texttt{ethutils} Git dependency are pinned. 
Required Solidity compiler releases are selected with \texttt{solc-select}, and compiler installation failure stops the pipeline. 
The verification command \texttt{scripts/verify\_rebuild.sh} first requires a clean tracked worktree, validates the input JSON, runs the repository-integrity checks, and then rebuilds correspondence validation, labels, location records, the integrated catalog, and metadata in their prescribed order.
It validates the rebuilt JSON and fails if any tracked output differs from the committed release.

\section*{Usage Notes}
For a compact set of correspondence-matched, non-refuted, localized contract cases, \path{cve.json} is the primary choice.
Use \path{data/index.csv} for coverage studies, library cases, refuted claims, or rejected correspondences.
Regarding detailed evidence and validation for artifact correspondence, labels, or vulnerability locations, refer to the directories \path{validation}, \path{labeling} and \path{localization}, respectively.

A correspondence \textit{match} must not be interpreted as an independent vulnerability finding. 
Likewise, labels characterize CVE claims and do not verify exploitability. 
Deployment addresses are intentionally absent when the source association is supported, but the deployment association is not.
Researchers evaluating detection tools may also include the refuted records, non-matches, library cases, and inherited entry points if useful.

Source artifacts are heterogeneous. Some are flattened verified sources, others are multi-file packages, and historical artifacts may depend on older compiler behavior. 
Compiler versions should be taken from the verified source in \path{data/etherscan} or provenance documentation rather than inferred from the current tool chain. 
Users must also observe third-party source licenses and notices.

\subsection*{Known limitations}
The dataset is based on records identified in one specific CVE snapshot; it is neither a complete inventory of smart contract vulnerabilities nor a guarantee that every relevant CVE record has been retrieved.
The dataset inherits omissions, errors, changing status, and uneven detail from CVE records and their linked references.
Supporting evidence as referenced by the CVE records varies by record and may later disappear or be revised, although retained evidence and provenance document what was used for this release.

Correspondence validation verifies sufficient agreement between CVE claims and the retained artifacts, but not the existence, severity, reachability, or exploitability of the claimed vulnerability.

A \texttt{non\_match} rejects the respective agreement rather than the existence of the upstream CVE claim.
A record is \emph{refuted} when we found convincing evidence that the CVE claim is a false positive.
Rule-based taxonomy mappings encode the reported claim and can inherit ambiguity from its wording; null mappings identify cases in which a more specific taxonomy assignment was not justified.
Similarly, a location indicates where the retained evidence places the claim.
It may contain multiple affected functions, statements, or reachable entry points and need not identify a unique minimal vulnerable statement.

The dataset inherits various biases from the CVE database.
It focuses on Ethereum, with only one record referring to a BNB Smart Chain contract.
The uneven number of records per year skews the distribution of Solidity compiler versions and of vulnerability types to versions 0.4.x and to integer bugs (cf.\ Table~\ref{tab:vuls-per-year}).

Finally, third-party rights determine which acquired materials can be redistributed and under what terms; users should consult the file \path{THIRD_PARTY_NOTICES.md} in the dataset.

\section*{Data Availability}
The dataset is archived on Zenodo~\cite{zenodo}, while GitHub hosts the maintained development repository~\cite{CVE-Smart-Contracts}.
The archive contains the integrated catalog, a complete index, selected CVE records, source and runtime artifacts, validation assessments, vulnerability labels, source locations, correspondence evidence, review decisions, schemas, and metadata described in the Data Records section.
Dataset-authored metadata, schemas, documentation, labels, annotations, and review decisions are licensed under CC BY 4.0.
Third-party source code, CVE records, explorer responses, runtime bytecode, and archived reference material remain under their original licenses, as detailed in \path{THIRD_PARTY_NOTICES.md}.

\section*{Code Availability}
The Zenodo archive and development repository~\cite{zenodo,CVE-Smart-Contracts} include the custom construction, validation, labeling, localization, integration, metadata, and rebuild-verification scripts.
Installation instructions, pinned dependencies, required Solidity compilers, execution order, and rebuild commands are documented in \path{README.md} and \path{REPRODUCIBILITY.md}.
Except where a file states otherwise, code written for the repository is licensed under the MIT License.

\section*{Author Contributions}
Monika di Angelo and Gernot Salzer contributed equally to conceptualization, methodology, software, validation, investigation, data curation, and writing.

\section*{Competing Interests}
The authors declare no competing interests.

\section*{Acknowledgements}

Thanks to Brian `Jericho' Martin and Kurt Seifried, former CVE board members, for patiently answering our questions.


\begin{thebibliography}{99}
\bibitem{CVE} CVE Program. \emph{CVEListV5: repository of CVE records.} \url{https://github.com/CVEProject/cvelistV5} (accessed 24 July 2026).
\bibitem{Wood} Wood, G. \emph{Ethereum: A secure decentralised generalised transaction ledger.} Ethereum Project \url{https://ethereum.github.io/yellowpaper/paper.pdf} (2025).
\bibitem{Solidity} Solidity Project. \emph{Introduction to smart contracts.} \url{https://docs.soliditylang.org/en/latest/introduction-to-smart-contracts.html} (accessed 12 August 2026).
\bibitem{EtherscanVerification} Etherscan. \emph{What is contract verification?} \url{https://docs.etherscan.io/contract-verification/whats-contract-verification} (accessed 29 July 2026).
\bibitem{cve_overview} CVE Program. \emph{Overview.} \url{https://www.cve.org/About/Overview} (accessed 24 July 2026).
\bibitem{msg00022} Seifried, K. \& Martin, B. CVE's for ``smart'' contracts, legal execution engines. \emph{CVE Board email archive} (2016). \url{https://www.cve.org/Resources/Media/Archives/OldWebsite/data/board/archives/2016-06/msg00022.html}
\bibitem{msg00004} Martin, B., Seifried, K. \& Meunier, P. Recent wave of smart contract vulnerabilities---out of scope? \emph{CVE Board email archive} (2018). \url{https://www.cve.org/Resources/Media/Archives/OldWebsite/data/board/archives/2018-07/msg00004.html}
\bibitem{msg00026} Coffin, C. CVE Board meeting summary---22 August 2018. \emph{CVE Board email archive} (2018). \url{https://www.cve.org/Resources/Media/Archives/OldWebsite/data/board/archives/2018-08/msg00026.html}
\bibitem{VeriSmart} So, S., Lee, M., Park, J., Lee, H. \& Oh, H. VeriSmart: A highly precise safety verifier for Ethereum smart contracts. In \emph{2020 IEEE Symposium on Security and Privacy}, 1678--1694 (IEEE, 2020). \url{https://doi.org/10.1109/SP40000.2020.00032}
\bibitem{VeriSmartGit} Software Analysis Lab. @ Korea University. \emph{VeriSmart-benchmarks} \url{https://github.com/kupl/VeriSmart-benchmarks/tree/master/benchmarks/cve}
\bibitem{SmartTest} So, S., Hong, S. \& Oh, H. SmarTest: Effectively hunting vulnerable transaction sequences in smart contracts through language model-guided symbolic execution. In \emph{30th USENIX Security Symposium}, 1361--1378 (USENIX Association, 2021). \url{https://www.usenix.org/system/files/sec21-so.pdf}
\bibitem{SmartFix} So, S. \& Oh, H. SmartFix: Fixing vulnerable smart contracts by accelerating generate-and-verify repair using statistical models. In \emph{31st ACM Joint European Software Engineering Conference and Symposium on the Foundations of Software Engineering}, 185--197 (ACM, 2023). \url{https://doi.org/10.1145/3611643.3616341}
\bibitem{SmartFixGit} Software Analysis Lab. @ Korea University. \emph{SmartFix-Artifact} \url{https://github.com/kupl/SmartFix-Artifact/tree/main/benchmarks/cve}
\bibitem{Etherscan} Etherscan. \emph{Etherscan API V2.} \url{https://docs.etherscan.io/api-reference/endpoint/getsourcecode} (accessed 28 August 2026).
\bibitem{taxonomy} Iuliano, G. \& Di Nucci, D. Smart contract vulnerability taxonomy. \emph{Journal of Systems and Software} (2026). \url{https://doi.org/10.1016/j.jss.2026.112788}
\bibitem{SWCregistry} Smart Contract Weakness Classification project. \emph{Smart Contract Weakness Classification Registry.} \url{https://swcregistry.io} (accessed 10 August 2026).
\bibitem{CWE} MITRE. \emph{Common Weakness Enumeration.} \url{https://cwe.mitre.org} (accessed 10 August 2026).
\bibitem{CVE-Smart-Contracts} di Angelo, M. \& Salzer, G. \emph{CVE Smart Contracts.} GitHub \url{https://github.com/smartbugs/CVE-Smart-Contracts/tree/ed62986909d00f7260e38178fe9a6cd5064c9ef6} (2026).
\bibitem{zenodo}  di Angelo, M. \& Salzer, G. \emph{CVE Smart Contracts.} Zenodo: \url{https://doi.org/10.5281/zenodo.22172881} (2026).
\end{thebibliography}
\end{document}